\documentclass[twocolumn]{aastex7}

\newcommand{\teff}{\ensuremath{T_{\rm eff}}}
\newcommand{\msun}{M\ensuremath{_\odot}}
\newcommand{\prot}{\ensuremath{P_{\rm rot}}}

\newcommand{\fspot}{$f_{\rm spot}$}
\graphicspath{{./}{Figures/}}
\submitjournal{ApJL}

\shorttitle{Radius Inflation and Magnetic Starspots}
\shortauthors{Cao \& Stassun}

\begin{document}

\title{The Relationship of Stellar Radius Inflation to Rotation and Magnetic Starspots at 10--670~Myr}

\author[0000-0002-8849-9816]{Lyra Cao}
\email{lyra.cao@vanderbilt.edu}
\affiliation{Department of Physics and Astronomy, Vanderbilt University, Nashville, TN 37235, USA}

\author[0000-0002-3481-9052]{Keivan G.\ Stassun}
\email{keivan.stassun@vanderbilt.edu}
\affiliation{Department of Physics and Astronomy, Vanderbilt University, Nashville, TN 37235, USA}

\begin{abstract}
Active, low-mass stars are widely observed to have radii that are larger than predicted by standard stellar models. Proposed mechanisms for this radius inflation generally involve stellar magnetism, either in the form of added pressure support in the outer layers and/or suppression of convection via starspots. We have assembled a large sample of 261 low-mass stars in the young clusters Upper Scorpius, $\alpha$~Persei, Pleiades, and Praesepe (spanning ages 10--670~Myr) for which the data exist to empirically measure the stellar radii, rotation periods, and starspot covering fractions. We find a clear, strong relationship between the degree of radius inflation and stellar rotation as represented by the Rossby number; this inflation-rotation relationship bears striking resemblance to canonical activity-rotation relationships, including both the so-called linear and saturated regimes. We also demonstrate here for the first time that the radius inflation depends directly on the starspot covering fraction. We furthermore find that the stars' effective temperatures decrease with decreasing Rossby number as well, and that this temperature suppression balances the radius inflation so as to preserve the stellar bolometric luminosity. These relationships are consistent across the age range sampled here, which spans from the pre--main-sequence to the zero-age main sequence. The favorable comparison of our findings to the predictions of modern starspot-based stellar evolution models suggests that, while rotation is clearly the underlying driver, magnetism may be the most likely direct cause of the radius inflation phenomenon. 
\end{abstract}


\section{Introduction} \label{sec:intro}
A consensus is emerging that some low-mass stars ($M \lesssim 1$~\msun) have larger radii than expected from standard stellar theory. This so-called ``radius inflation'' effect is typically of order 10--15\% and is correlated with a roughly 5--10\% lower effective temperature than predicted (``temperature suppression"). Radius inflation has been observed in numerous studies using a variety of observational techniques including eclipsing binary analysis \citep[e.g.,][]{1997AJ....114.1195P,2002ApJ...567.1140T,2005ApJ...631.1120L,Stassun:2022}, statistical projected radii \citep[e.g.,][]{2016A&A...586A..52J,2018MNRAS.476.3245J}, and spectral energy distributions \citep[SEDs; e.g.,][]{2017AJ....153..101S,2019ApJ...879...39J,2023ApJ...951...90W}. 

Although the precise mechanism is still debated, strong magnetic activity seems to play a role either through the direct inhibition of convective energy transport, the influence of large starspots on the photospheric pressure and temperature, or a combination of both effects \citep[e.g.,][]{2001ApJ...559..353M,2007A&A...472L..17C,2010ApJ...723.1599M,2013ApJ...779..183F,2014ApJ...789...53F,2014MNRAS.445.4306J,2014MNRAS.441.2111J,2014ApJ...790...72S,2015MNRAS.449.4131S,2015ApJ...807..174S}. This conclusion is based on observed correlations between radius inflation and proxies of magnetic activity such as H$\alpha$ emission, X-rays, photometric variability, and rotation rate \citep[e.g.,][]{2005ApJ...631.1120L,2012ApJ...756...47S,2017AJ....153..101S,2025ApJ...982..114M}.

Prior studies have suggested that the radius inflation mechanism may be connected to the as-yet unclear physics of magnetic saturation. \citet{2017AJ....153..101S} and \citet{2019ApJ...879...39J} investigated radius inflation among K-dwarfs in the Pleiades and Hyades open clusters, respectively, using empirically determined radii and rotation period measurements.
They found that for stars rotating with a period slower than 2~days, corresponding to a Rossby number\footnote{The Rossby number is defined as the ratio of the rotation period to the convective overturn timescale \citep[e.g.][]{1984ApJ...279..763N}.} ($Ro$) greater than $\sim$0.1 in the mass range they studied ($\sim$0.7--0.9~\msun), standard stellar models predicted the radii well. However, stars rotating faster than 2~days ($Ro\lesssim 0.1$) were on average $\sim$$12$\% larger than predicted.
This is an interesting value of $Ro$, as numerous other studies have found that the correlation between rotation rate and magnetic proxies saturates at approximately this value \citep[e.g.,][]{2011ApJ...743...48W, 2022A&A...662A..41R, 2022MNRAS.517.2165C}. 

\citet{2012ApJ...756...47S} developed empirical relationships between radius inflation, temperature suppression, and H$\alpha$ luminosity as an activity tracer, but the available samples were small and heterogeneous, and therefore the signal was noisy.
More generally, while previous studies investigating radius inflation find that rapid rotators are systematically more active than slower rotators, it has been challenging to interpret radius inflation directly as an activity-dependent signal.

The radius inflation scenario has not yet been tested at ages younger than the Pleiades, which is a major source of uncertainty for realistic models underlying the processes and timescales of star and planet formation. There is reason to believe this effect is significant in young stars. Some of these stars appear superlatively active \citep{2017ApJ...836..200G, 2022MNRAS.517.2165C}. If magnetic activity is primarily responsible for the radius inflation phenomenon, then the inflation signals in stars should be related to their observed activity level---a hypothesis testable in stars spanning a wide range in age and origin.

The radius inflation effect is with respect to some standard,
namely, standard (non-magnetic) stellar models; stellar models that incorporate the effects of magnetism \citep[e.g.,][]{2014ApJ...789...53F, 2015ApJ...807..174S, 2020ApJ...891...29S} are candidate solutions. In star-forming regions, magnetic model isochrones appear to resolve other observational tensions, such as reducing discrepancies in isochronal ages and radii for stars formed in the same environment---with the potential to shift age estimates by as much as a factor of two \citep{2016A&A...593A..99F, 2020ApJ...891...29S, 2022ApJ...924...84C}. Magnetic stellar models have the potential to more accurately predict the observed radii of stars, especially in young clusters, but a lack of available magnetic activity data has so far made these comparisons unfeasible.

Recent advances in the modeling and analysis of stellar spectroscopic data have made it possible to study magnetic fields in detail by measuring starspot covering fractions and estimating temperatures of the inhomogeneous spotted surfaces of active stars \citep{2022MNRAS.517.2165C}, applicable to the large number of stellar clusters observed by APOGEE DR17 \citep{2021AJ....162..302B}. Together with the broad availability of precise photometric SEDs and fitting methods \citep[e.g.,][]{2016AJ....152..180S,2017AJ....153..101S,2019ApJ...879...39J} and accurate parallaxes and astrometry from {\it Gaia}---which make it possible to empirically determine radii in conjunction with estimates of stellar extinction, cluster membership, and binarity---and rotational studies with {\it Kepler\/} and {\it TESS\/} that have made it possible to perform this kind of dynamo analysis on a large number of stars \citep[e.g.,][]{2016AJ....152..113R, 2018AJ....155..196R, 2023AJ....166...14B, 2017ApJ...839...92R}, we are now able to directly study empirical relationships between radius inflation, temperature suppression, rotation, and stellar magnetism in an unprecedentedly detail-rich and self-consistent fashion.

This paper presents such an analysis for a sample of 261 cool stars belonging to open clusters with ages spanning 10--670~Myr. Section~\ref{sec:methods} summarizes the study sample, data used, and methods applied to their analysis. Section~\ref{sec:results} presents the main results, in the form of statistically meaningful, empirical relationships between radius inflation, temperature suppression, stellar rotation, and magnetic starspot properties. We conclude with a discussion and summary of these results in Section~\ref{sec:disc}.

\begin{figure*}[!ht]
    \centering
    \includegraphics[width=0.75\linewidth,trim=0 0 0 0,clip]{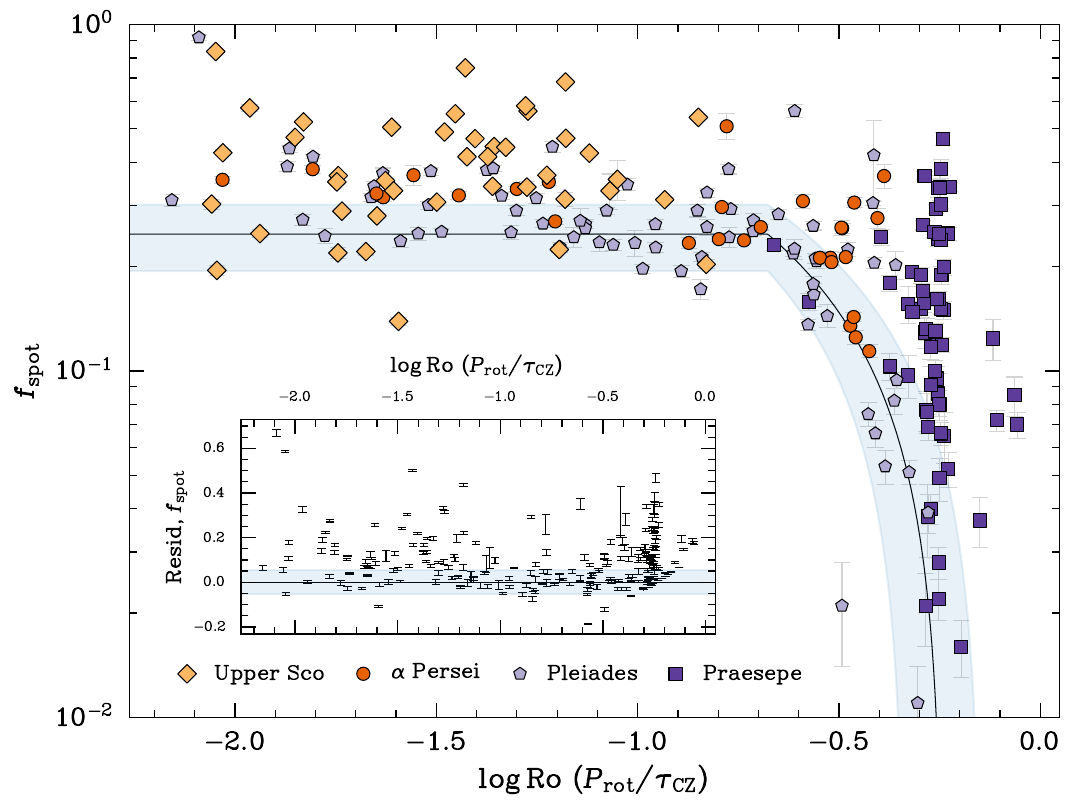}
    \caption{Magnetic starspot filling fractions as a function of the rotational Rossby number parameter for stars in our study sample. Stars are colored and assigned a marker according to their host cluster. For comparison, the light blue swathe represents the fitted relation found by \citet{2022MNRAS.517.2165C} in their study of the Pleiades, which appears to trace the lower envelope of the spot--Rossby relation for our larger sample of open clusters with ages spanning 10--670~Myr.}
    \label{fig:fspot_vs_Ro}
\end{figure*}

\section{Data \& Methods} \label{sec:methods}

\subsection{Study Sample}\label{sec:sample}
To track the evolution of the radius inflation signal, we begin with stars from well-known open clusters with ages in the range 10--670 Myr: Upper~Sco \citep[$\sim$10~Myr;][]{2016A&A...593A..99F}, $\alpha$~Per \citep[$\sim$80~Myr;][]{2023AJ....166...14B, 2022A&A...664A..70G}, the Pleiades \citep[$\sim$112~Myr;][]{2015ApJ...813..108D}, and Praesepe \citep[$\sim$670~Myr;][]{2019ApJ...879..100D}. We derive initial membership lists from the main rotation-period studies for Upper~Sco \citep[][]{2018AJ....155..196R}, $\alpha$~Per \citep[][]{2023AJ....166...14B}, the Pleiades \citep[][]{2016AJ....152..113R}, and Praesepe \citep[][]{2017ApJ...839...92R}. We then crossmatched these datasets with LEOPARD \citep[][Cao et al.\ \emph{(in prep)}]{2022MNRAS.517.2165C, 2023ApJ...951L..49C}, a spectroscopic starspot catalog derived from APOGEE DR17 data \citep{2022ApJS..259...35A}. This led to an initial catalog of 106 stars in Upper Sco, 86 stars in $\alpha$~Per, 256 stars in the Pleiades, and 133 stars in Praesepe.

To ensure that cluster members are free from binary or blending contamination---a potential source of concern when measuring stellar radii and starspot signals---we describe our binary rejection technique in Appendix~\ref{sec:binarystars}. 

For each star, we obtained where possible the stellar radius via the ``pseudo-interferometry" method described in Section~\ref{sec:cluster}, applied to the broadband optical-IR photometric SED, and excluded stars for which our solution was too discrepant ($\chi_{\mathrm{SED}}^2 > 10$). In our youngest cluster, Upper~Sco, the presence of accretion may manifest in the stars' spectra as veiling, an effect most dramatic at UV and optical wavelengths \citep{1990ApJ...363..654B} which could directly contaminate the radii determined from the SED analysis, or reliable two-temperature spectroscopic fits. Therefore, we exclude stars that have color excesses in the mid-infrared consistent with accreting disks \citep[i.e., excluding stars with WISE colors $W1 - W2 > 0.3$, following prescriptions from][]{2015AJ....150..100K, 2019MNRAS.486.1907J, 2022ApJ...924...84C}.

In total, our study sample includes 47 stars in Upper~Sco, 36 stars in $\alpha$~Per, 86 stars in the Pleiades, and 92 stars in Praesepe. We graphically introduce the full sample in Figure~\ref{fig:fspot_vs_Ro}, demonstrating the expected mean correlation between starspot filling fractions (\fspot) and Rossby number ($Ro$) from \citet[][]{2022MNRAS.517.2165C}.

\subsection{Determination of Stellar Properties: Radii, Starspots, and Rotational Dynamos} \label{sec:cluster}

To obtain stellar radii measurements, we perform fits to the stars' broadband optical-IR SEDs, simultaneously with dereddening, again following the methodology of \citet{2016AJ....152..180S,2017AJ....153..136S,2018AJ....155...22S} as was done in our previous radius-inflation studies of the Pleiades and Hyades \citep{2017AJ....153..101S,2019ApJ...879...39J}.

The underlying SED functions are drawn from the NextGen grid of atmosphere models. In this way, we obtain 
the empirically measured bolometric flux at Earth ($F_{\mathrm{bol}}$), which is then converted to an angular diameter, $\theta_{\mathrm{bol}}$, via the Stefan-Boltzmann relation. For consistency in our analysis, we adopt the best-fit spectroscopic {\teff} from the flux-weighted two-temperature starspot fits described above. Finally, applying the {\it Gaia\/} DR3 parallax to $\theta_{\mathrm{bol}}$ yields the stellar radius. Errors were estimated through direct numerical sampling, assuming errors in the stellar \teff\ and $F_{\mathrm{bol}}$ measurements were normally distributed and independent.

Using the two-temperature spectroscopic fitting technique from \citet{2022MNRAS.517.2165C}, we obtain starspot filling fractions jointly with six other stellar parameters (microturbulence, $v \sin i$, metallicity, \teff, surface gravity, and starspot temperature contrast) on the high-resolution (R${\sim}$22,500) near-infrared H-band spectra from APOGEE.

For the analysis of stellar dynamo evolution, we convert rotation-period (\prot) measurements from the literature (see Section~\ref{sec:sample}) to rotational Rossby number ($Ro \equiv \prot/\tau_{\mathrm{CZ}}$). To do this, we require estimates of the convective turnover timescale ($\tau_{\mathrm{CZ}}$), which we estimate theoretically from SPOTS models \citep{2020ApJ...891...29S}, using observed $f_{\mathrm{spot}}$ to lookup appropriate convective overturn timescales.

\subsection{Spotted and Rotating Stellar Models} \label{sec:magevol}

\citet{2020ApJ...891...29S} released a set of non-rotating stellar isochrone models (SPOTS) that include the impact of starspots on surface convection \citep{2015ApJ...807..174S}. This prescription involves (1) suppression of convection due to surface starspot coverage, (2) stabilization of convection through a magnetic modification to the Schwarzschild criterion, and (3) addition of a magnetic pressure term at the surface. 

The SPOTS models include \fspot\ as a primary, independent parameter. 
To connect the \fspot\ parameter in the SPOTS models to the stars' rotational properties, in our analysis we use the rotationally evolved ROTEVOL models, which are based on the SPOTS models but calibrated to the stellar rotation distributions in the Pleiades and Praesepe from \citet{2023ApJ...951L..49C}, in a procedure that self-consistently matches rotation periods, metallicities, and measured \fspot\ values of stars in the clusters.

We define all of our measured radius inflation signals relative to the base case (non-rotating SPOTS isochrones with \fspot\ set to zero), using the fiducial ages reported in the literature for each cluster. The spotted, rotating ROTEVOL calibrated isochrones are then the test case from stellar theory, and can be compared directly to predictions from the base case.

\section{Results \& Analysis} \label{sec:results}
\subsection{The Radius Inflation---Rossby Number Relation}

\begin{figure*}[!ht]
    \centering
    \includegraphics[width=0.75\linewidth,trim=0 0 0 0,clip]{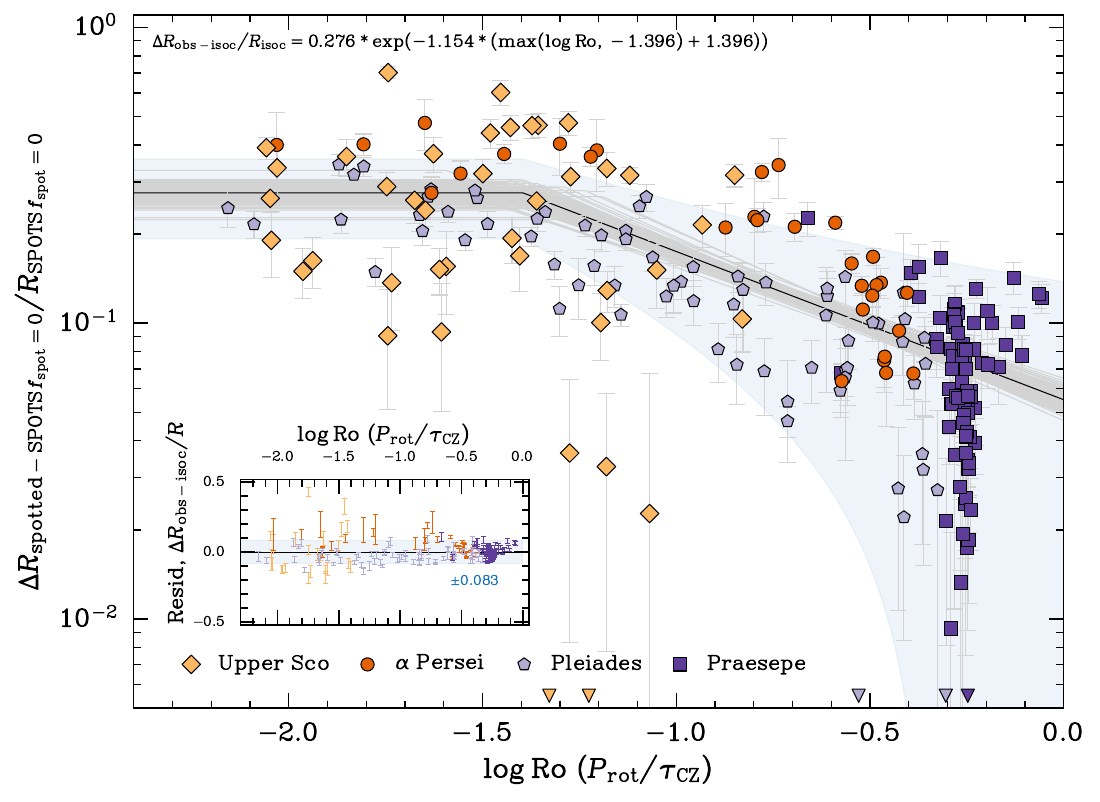}
    \includegraphics[width=0.75\linewidth,trim=0 0 0 0,clip]{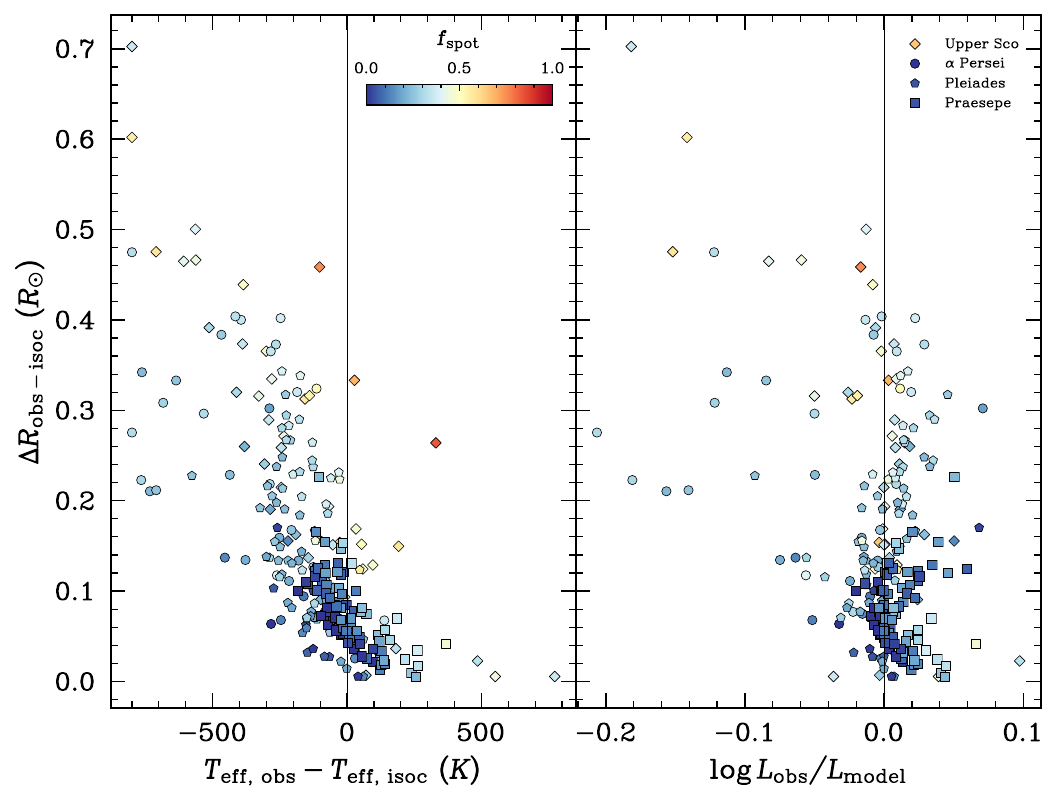}
    \caption{{\it (Top:)} Degree of (fractional) radius inflation for our study sample, determined in relation to standard, non-magnetic, unspotted stellar isochrone models, as a function of the rotational Rossby number. Stars are color-coded and assigned a marker by their cluster membership. The black line is the fiducial ``best fit'' for the power law relation, the gray lines indicate alternate sampled solutions of the MCMC chain (representing the uncertainty of the solution), with the blue band representing the standard deviation of points about the best fit line. Points which fall off the edge of the plot limits are denoted by triangle symbols, but are still included in the fit.
    {\it (Bottom:)} Observed radius inflation as a function of temperature suppression (left) and luminosity offsets from SPOTS isochrones (right). Temperature suppression is observed to correlate with radius inflation in a manner that preserves the luminosity.}
    \label{fig:DelR_frac_vs_Ro}
\end{figure*}

Figure~\ref{fig:DelR_frac_vs_Ro} (top) presents the basic result of this study: the degree of radius inflation across our study sample is a strong function of Rossby number, strongly suggesting a dynamo origin for the radius inflation phenomenon that operates in clusters over a span of ages. The resulting dataset for our sample is in Table \ref{tab:clusterdata}, and the MCMC procedure used to determine the best-fit relationship is described in Appendix~\ref{appendix:mcmc}. 

We observe across the entire study sample a clear relationship that mimics the standard ``rotation-activity relation" with both linear and saturated regimes. Prior observations in individual clusters had shown inflated radii for rapid rotators in individual clusters to be correlated with lithium and rotation \citep{2017AJ....153..101S, 2019ApJ...879...39J}. However, the break between the two regimes at $\log Ro \approx -1.4$ appears to be at a lower saturation point than reported for other proxies in the literature. The break in the inflation--rotation relationship appears universal, as it is not inferred simply by looking at clusters which sample the saturated and unsaturated domains separately; rather, the break in the inflation--rotation relation is seen in clusters with stars spanning the saturated and unsaturated domains.

The radius inflation signal in the Rossby plane shows a striking similarity with the saturation pattern of other activity proxies. The signal is seen in stars which also demonstrate a rotational saturation in the starspot filling fractions (as seen in Figure~\ref{fig:fspot_vs_Ro}). It is challenging to directly compare with the literature, because the definition of the convective overturn timescale can differ between studies (see Figure 2 \& Appendix Figure A1 in \citet{2022MNRAS.517.2165C}). For our own sample, we observe a different critical Rossby number associated with radius inflation as opposed to starspot coverage.

The vertical offsets of each cluster in the inflation--Rossby relation can be affected by the assumed age of the cluster: a larger cluster age skews inflation estimates larger, as isochrones contract toward the ZAMS. The tendency for nearly all of the $\alpha$~Per stars to lie above the fit and for the Pleiads to lie below is likely related to their assigned ages. As noted in \citet{2023AJ....166...14B}, recent estimates for the age of $\alpha$~Per appear to prefer $\sim$80~Myr: from the lithium depletion boundary (LDB) age ($79^{+1.5}_{-2.3}$~Myr) \citep{2022A&A...664A..70G} and the gyrochronal age ($86\pm16$~Myr) \citep{2023AJ....166...14B}. We assumed a LDB age for the Pleiades ($112\pm5$~Myr) \citep{2015ApJ...813..108D}.
From the radius inflation and starspot analysis we perform in this paper, we find that a larger age difference than we currently assume (32 Myr) between the two clusters would be favored if the radius inflation signal were to be interpreted as universal.

\subsection{Radius Inflation in Relation to Temperature and Luminosity}\label{sec:tempsuppression}

Previous investigations observed that radius inflation in cool stars is often accompanied by a reduction of surface temperature, and that the combined effect of this temperature suppression with radius inflation is such that the bolometric luminosity of the star is preserved. For example, in their study of low-mass eclipsing binaries, \citet{2012ApJ...756...47S} defined empirical relations that linked radius inflation and temperature suppression, independently from one another, to H$\alpha$ luminosity, finding that $\Delta R \sim \Delta\teff^{-2}$, as would be expected from the Stefan-Boltzmann relation if the luminosity remains constant. 
Figure~\ref{fig:DelR_frac_vs_Ro} (bottom) explores this question for our study sample, where we plot the observed radius inflation signal against the difference between observed and inferred temperature for a given model (left), and against observed and inferred luminosity (right). We observe that systematics associated with our radius inflation signal appear to be driven by the temperature suppression effect.

One explanation for these relationships between radius inflation, temperature suppression, and luminosity originates from the fact that although the stellar luminosity is generated in the core---far from the direct effects of the surface magnetic field---it can be affected by the action of fields on surface convection zones. The presence of dark magnetic starspots on the stellar surface is thought to affect observables depending on the depth of the convection zone and the degree to which they are spotted \citep{1986A&A...166..167S}. If the field indeed serves as the driver of radius inflation (see Section~\ref{sec:driver}), it could do so by retarding the emergent flux through reduced convective efficiency and/or dark starspots, reducing \teff\ in the process; then it is only through an increased surface area (inreased radius) that the stellar luminosity from the core can be emitted at the surface. 
See also \citet[][and references therein]{Morales:2010} for further discussion of these ideas.

\subsection{Stellar Magnetism as a Driver of Radius Inflation}\label{sec:driver}
\begin{figure*}
    \centering
    \includegraphics[width=0.8\linewidth,trim=0 0 0 0,clip]{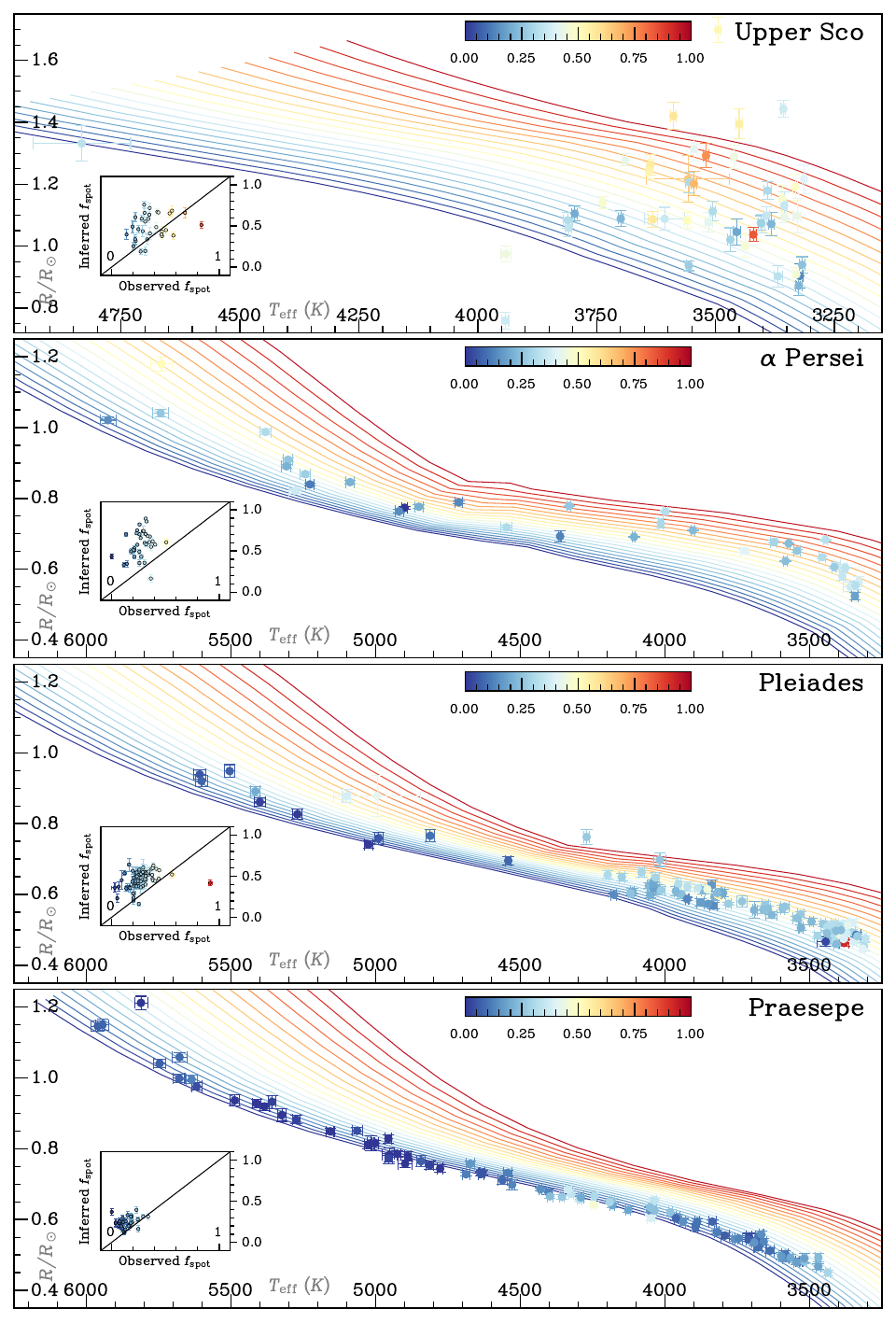}
    \caption{Comparisons of radii and temperatures recovered for individual stars, to those predicted by the SPOTS models, as a function of \fspot. Markers are colored by their $f_{\mathrm{spot}}$  
    and lines are also colored by the corresponding $f_{\mathrm{spot}}$  theoretical starspot isochrones from SPOTS. Inset plots relate the observed $f_{\mathrm{spot}}$ recovered from the spectra (x-axis) to the spot filling fraction inferred from the isochrones using the measured radii (y-axis).}
    \label{fig:spots_comp}
\end{figure*}

If magnetic starspots are directly implicated in the radius inflation scenario, we might expect magnetic stellar evolutionary models---which explicitly attempt to account for the physical effects of starspots---to mitigate radius and temperature discrepancies in the pre-main sequence. To test this, we represent in Figure~\ref{fig:spots_comp} the rotating and spotted stellar evolutionary models from Section~\ref{sec:magevol} along with the stars from our study sample. We plotted in the inset figures the observed starspot filling factors, and infer a theoretical starspot filling factor from the isochrones by interpolating a starspot value from SPOTS on the \teff--$R/$R$_\odot$ plane.

We find that stars' observed radii and temperatures are generally well described in each cluster by magnetic stellar isochrones, shown by the tendency for points which depart from the $f_{\mathrm{spot}}=0$ isochrone to be more spotted.
Across multiple cluster sequences, this suggests that radius anomalies seen in single stars can be described in a systematic fashion by starspots.
Starspot models appear to underpredict the radius inflation effect for active stars by about 0.27~dex, suggesting that additional effects due to stellar magnetism may be relevant and need to be incorporated in stellar evolution models (see discussion in Section \ref{sec:disc}).

\begin{figure*}[!ht]
    \centering
    \includegraphics[width=0.8\linewidth,trim=0 0 0 0,clip]{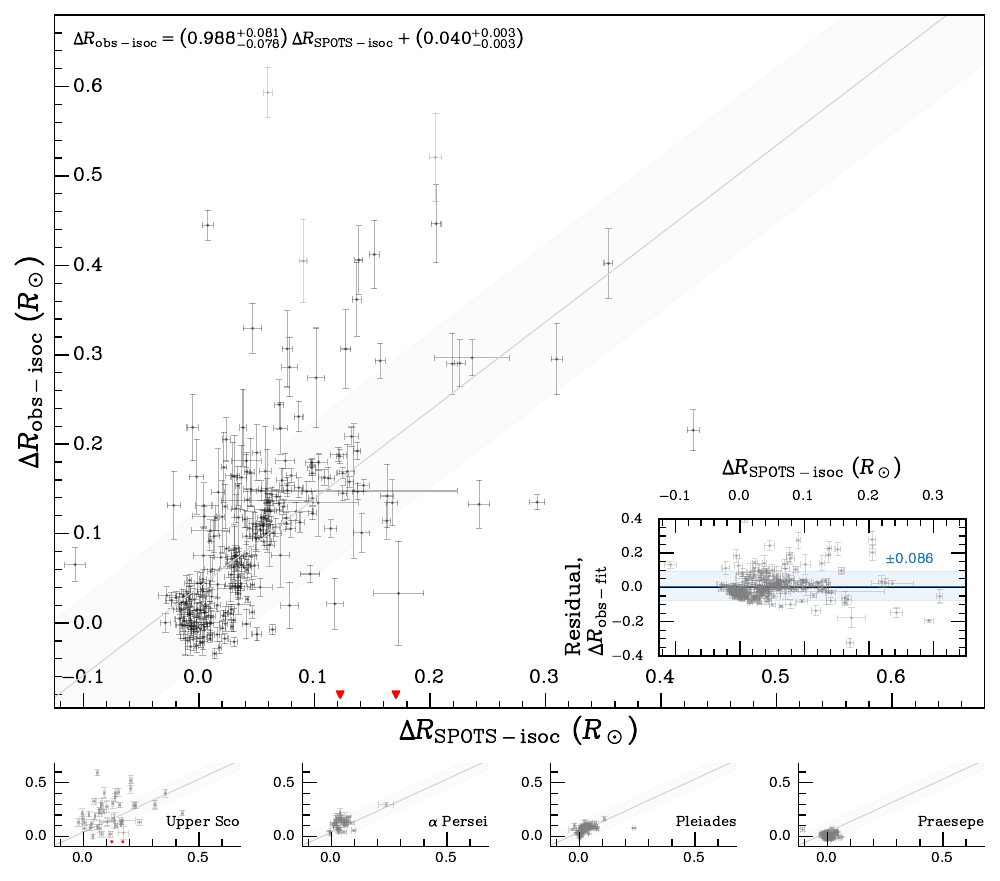}
    \caption{Direct comparison of the degree of observed radius inflation and predictions from the magnetic SPOTS models. Points are color-coded by starspot filling factor, and the marker shape corresponds to the cluster. Inset plot shows the residual in the observed radius inflation parameter relative to model predictions. Bottom plots show the mean relation with just the individual stars of each cluster. The robust 1:1 correlation between observed inflation values and the predicted values suggests that the magnetic stellar evolution model makes consistent radii estimates of inflated stars. Two points from Upper Sco (indicated by red upper limits) are not shown on the diagram, but they are still included in the fit.}
    \label{fig:DelR_obs_vs_spots}
\end{figure*}

Finally, we directly test whether our measured stellar radii are consistent with those predicted by magnetic stellar models. For a star with a known \teff, age, and starspot filling factor, we compared the SPOTS-predicted radii to those that we observed in Figure~\ref{fig:DelR_obs_vs_spots}.
The correlation between the radii observed from our sample and the radii predicted by the magnetic stellar models is strong and clear, with a Kendall's $\tau$ statistic of $0.492$, corresponding to a statistical confidence of $>$99.9\% in rejecting the null hypothesis of no correlation. 

Overall, the slope of the correlation for all stars, based on a MCMC fit, is consistent with unity ($0.99\pm0.08$). Again, the observed radius inflation appears to be systematically larger than the SPOTS-predicted radius inflation, by $\sim$$0.040\pm0.003$~R$_\odot$ (the standard error, shown in parentheses in the figure). Despite the fact that the observed inflation signal appears to be underestimated by a factor of 2 in Figure~\ref{fig:spots_comp}, the good agreement between model radii even at high levels of inflation in Figure~\ref{fig:DelR_obs_vs_spots} suggests that starspot models can already provide consistent radius inferences to stars in the pre-main sequence by accounting for stellar magnetism.

\section{Discussion \& Conclusions} \label{sec:disc}
We demonstrate a clear radius inflation signal across multiple clusters spanning ages from the end of the T~Tauri phase to the zero-age main sequence. The consistent inflation signal in pre-main sequence stars suggests that the level of radius inflation that we see is a universal function of the stellar activity level, rather than a property of the individual clusters. The departure of these stars from standard isochrones follows the same pattern as a standard activity--Rossby relation, implying a magnetic origin to the inflation scenario.

Other lines of evidence also indicate the essential role of starspots and magnetism in young stars for many stellar observables. \citet{2025MNRAS.tmp..654J} found that a model where stars differentially burn lithium according to star-by-star variations in spottedness appears to better explain observed cluster spreads in lithium abundances for pre-main sequence stars as early as 10~Myr. As radius inflation is also included in those models, which depress lithium depletion by cooling the base of the convection zone, this indicates that the mechanism by which depletion and radius inflation occur may be linked to an underlying expression of magnetism on pre-main sequence stars.

Activity-level effects in stars appear to affect stars by as much as $\sim$25\% (relative to standard isochrones) in this work, an effect which will need to be accounted for in active stars: young stars and low-mass stars. By ignoring activity effects in stellar characterization, stellar radii---and any quantities that depend on it, such as exoplanetary radii---can be affected by a strong activity-dependent bias. As a result, directly interpretable stellar magnetic proxies such as magnetic field strength and starspot filling factor are likely to be essential in delivering on the promise of precision stellar astrophysics.

The dispersion in starspot filling factors at constant Rossby number, as seen in Figure \ref{fig:fspot_vs_Ro}, appears larger than expected from the reported spectroscopic measurement errors. However, this variation has potential explanations---first, stars undergo magnetic cycles in their magnetic fields and activity levels \citep{2023MNRAS.525..455D, 2023SSRv..219...54J, 2018ApJ...855...75R}, which can naturally produce a spread in starspot coverage. Stars also exhibit a range of magnetic field configurations, including complex or dipolar field geometries with varying degrees of axisymmetry (for a review, see \citet{2009ARA&A..47..333D}); anisotropies in the surface geometry of starspots with respect to latitude (such as predominantly polar or equatorial spots) can also produce spreads in spot coverage as the stellar inclination varies.
Binarity is also a factor, but in open clusters we expect the binary contamination rate after our rejection procedure to be low (see Appendix \ref{sec:binarystars})---therefore, we expect dominant contributions to observed spreads to be from activity cycles, field geometry, and inclination effects.

The idea that the radius inflation saturation level is at a very small Rossby number, as we see in this work, is also very interesting. \citet{2022MNRAS.517.2165C} suggested in various fits (see their Figures 11 \& 12) that the saturation levels of different proxies appear to saturate at different critical Rossby numbers for the same stars, an effect that was also seen by \citet{2021ApJ...916...66B} for two starspot variability proxies. We continue to suggest that the point of Rossby saturation ($\mathrm{Ro}_{\mathrm{sat}}$) may be physically interesting, since the causative mechanism of the radius inflation may be anchored to a mechanism which has a characteristic length and pressure scale in stars which might be probed by this saturation behavior. Further careful exploration of how activity indices vary with Rossby number is needed to explore these scenarios.

The indication that the predicted degree of radius inflation is lower in the SPOTS models than observed suggests that more sophisticated treatments of stellar magnetism, and its interaction with other relevant physics like rotation and convection, may be necessary to improve consistency. The two-temperature inhomogeneous surface model of \citet{2020ApJ...891...29S} is simplistic, without considering other internal magnetic effects which may be relevant \citep[e.g.][]{2013ApJ...779..183F}. Alternatively, the radius inflation signal may be to some degree overestimated in our sample due to incomplete binary rejection. However, this possibility seems unlikely due to the persistence of this effect across likely single stars.

We find, despite these caveats, that the SPOTS models can consistently predict radii even for highly inflated stars, with a small radius offset. The inclusion of magnetism in stellar evolutionary models, such as with the SPOTS models, appears to explain much of the observed radius inflation in the pre-main sequence. The ability of magnetic stellar models to predict stellar radii of active low mass stars and active young stars is a strong argument for the inclusion of such models in stellar characterization workflows, towards the goal of precision stellar astrophysics; it also motivates the idea that a large catalog of physically interpretable magnetic proxies, such as surface field strength or starspot filling factor, may be key to characterizing young, low mass, or active stars.

Finally, we argue that isochronal discrepancies that many have reported in the literature in pre-main sequence clusters and in low-mass stars are likely primarily due to stellar magnetism, rather than some other artifact of the data or models. The systematic departures of isochrones in the clusters explored in this work can be largely explained as the result of starspots, using our own measured spectroscopic starspot filling factors. We find that discrepancies between stellar radii and theoretical predictions by standard stellar models appear to be driven by activity; the observed differences in the shapes of non-magnetic stellar isochrones and observed cluster sequences appear to be the manifestations of stellar magnetism. Thus, the path forward to reconciling issues with stellar evolutionary models throughout the pre-main sequence seems to be linked to acquiring a better understanding of the evolution of stellar magnetism.

\begin{acknowledgments}
We acknowledge funding support through the Vanderbilt Initiative in Data-intensive Astrophysics (VIDA). We acknowledge funding support through NASA Grant 80NSSC24K0622.

Funding for the Sloan Digital Sky Survey IV has been provided by the Alfred P. Sloan Foundation, the U.S. Department of Energy Office of Science, and the Participating Institutions. SDSS acknowledges support and resources from the Center for High-Performance Computing at the University of Utah. The SDSS web site is www.sdss4.org.

SDSS is managed by the Astrophysical Research Consortium for the Participating Institutions of the SDSS Collaboration including the Brazilian Participation Group, the Carnegie Institution for Science, Carnegie Mellon University, Center for Astrophysics | Harvard \& Smithsonian (CfA), the Chilean Participation Group, the French Participation Group, Instituto de Astrofísica de Canarias, The Johns Hopkins University, Kavli Institute for the Physics and Mathematics of the Universe (IPMU) / University of Tokyo, the Korean Participation Group, Lawrence Berkeley National Laboratory, Leibniz Institut für Astrophysik Potsdam (AIP), Max-Planck-Institut für Astronomie (MPIA Heidelberg), Max-Planck-Institut für Astrophysik (MPA Garching), Max-Planck-Institut für Extraterrestrische Physik (MPE), National Astronomical Observatories of China, New Mexico State University, New York University, University of Notre Dame, Observatório Nacional / MCTI, The Ohio State University, Pennsylvania State University, Shanghai Astronomical Observatory, United Kingdom Participation Group, Universidad Nacional Autónoma de México, University of Arizona, University of Colorado Boulder, University of Oxford, University of Portsmouth, University of Utah, University of Virginia, University of Washington, University of Wisconsin, Vanderbilt University, and Yale University.

This manuscript uses the \texttt{smplotlib} style \citep{jiaxuan_li_2023_8126529}.
\end{acknowledgments}

\bibliography{main}{}
\bibliographystyle{aasjournal}

\appendix
\renewcommand\thefigure{A.\arabic{figure}}
\setcounter{figure}{0}
\section{Binary Stars} \label{sec:binarystars}

Since binary stars can complicate the interpretation of the two-temperature spectroscopic fit \citep{2022MNRAS.517.2165C}, affect the SED fit \citep{2016AJ....152..180S}, and potentially affect the radius inflation signal, we elect to remove them in our cluster analysis. We adopt a similar approach to \citet{2022MNRAS.517.2165C} in rejecting binaries, by relying on the {\it Gaia} Re-normalized Unit Weight Error (\texttt{RUWE}) parameter, {\it Gaia} radial velocity variability flags, and multiple source detections, as well as the APOGEE \texttt{VSCATTER} parameter, which also indicates variations in radial velocities between APOGEE visits \citep{2022ApJS..259...35A}. We elected not to use a photometric binary cut since a spread in the HR Diagram for Upper Sco can be due to reasons other than binarity, such as an intrinsic age spread. In each of these parameters, we flag stars if they might be candidate binaries according to the criteria set in Sections 2.3.2--2.3.4 in \citet{2022MNRAS.517.2165C}.

\begin{figure*}[!ht]
    \centering
    \includegraphics[width=0.7\linewidth,trim=0 0 0 0,clip]{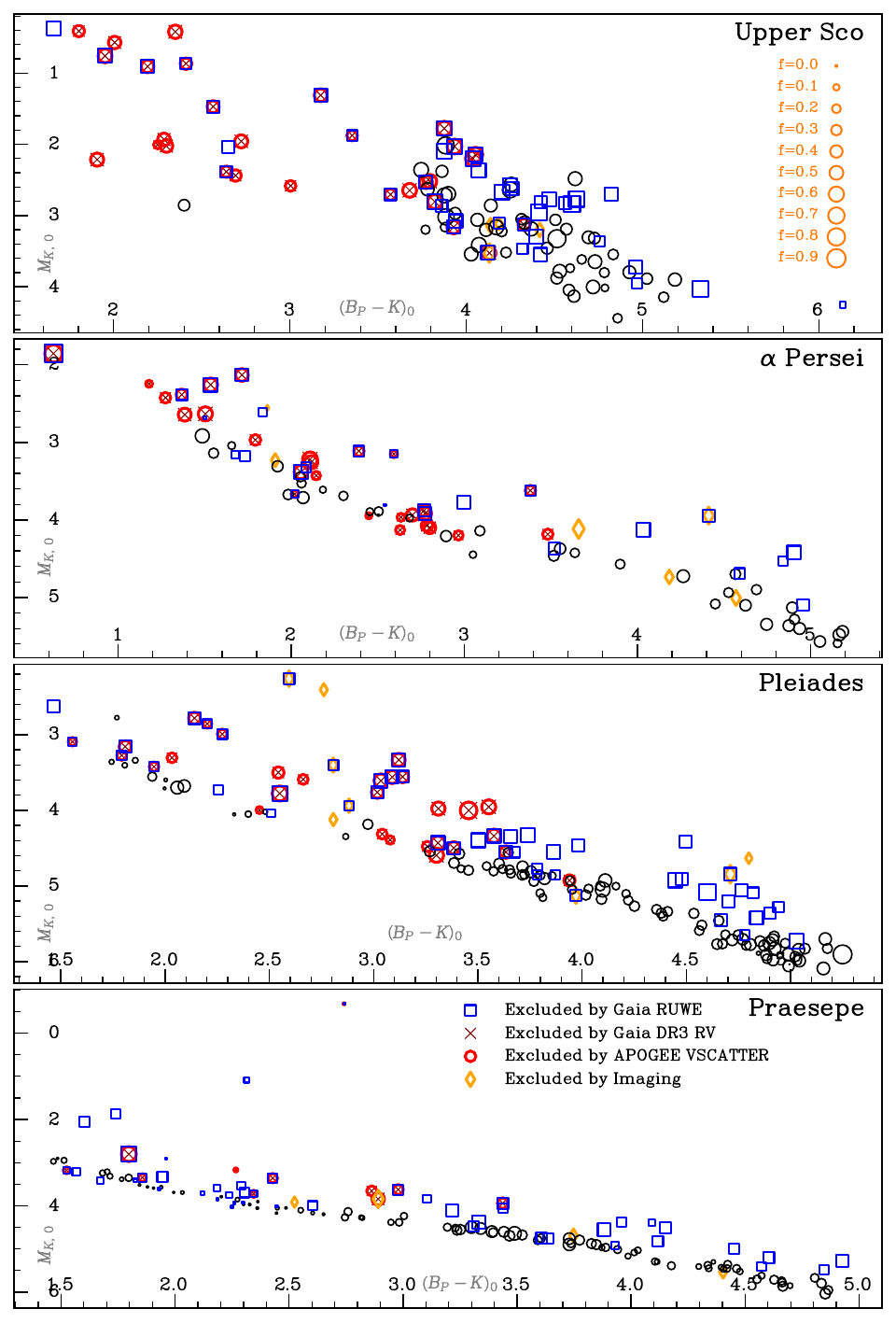}
    \label{fig:cluster_cmd}
    \caption{Binary rejection procedure on the clusters included in this analysis. Marker sizes correspond to the measured starspot filling fraction, and are shown overlapping in the event that individual stars are excluded due to one or more factors.}
\end{figure*}

In Figure~\ref{fig:cluster_cmd}, we show the results from our binary rejection technique. Stars are represented with a black circle if they are not excluded by any binary rejection criteria, or they are shown with some overlapping combination of colored symbols if they match one or more of the binary rejection criteria. A star is represented with a blue square marker if its \texttt{RUWE} parameter is greater than 1.2, a criterion which is commonly used to flag possible unresolved binarity \citep{2021ApJ...907L..33S, 2023A&A...674A..32B}. A maroon ``X'' is applied if a star exceeds the variability criteria in the {\it Gaia} RV variability indices, as identified in \citet{2023A&A...674A...5K}, specifically if \verb|rv_chisq_pvalue|$\leq 0.01$ \& \verb|rv_renormalized_gof|$>4$ \& \verb|rv_nb_transits|$\geq 10$ \& $3900 \leq$\verb|rv_template_teff|$\leq 8000$. A red circle is shown in the case that a star has \texttt{VSCATTER} $> 1$ km/s, which is seen in those stars with repeat visits and large observed RV variability. Finally, an orange diamond is plotted for a star if there is a neighboring source within 3 arcsec contributing more than 1 percent of the flux in {\it Gaia} $G_{BP}$ or $G_{RP}$, an indicator that an additional source could be visibly contaminating the observed spectrum. Following \citet{2022MNRAS.517.2165C}, we assume a $\gtrsim$85\% binary rejection completeness for our sample.

The dispersion in observed starspot filling factors seen in Figure~\ref{fig:fspot_vs_Ro} may partially be due to missed binaries, as the flux contribution by the companion can be misinterpreted as an activity signal and bias the measurement toward higher spot coverage. The flux contribution by an unresolved companion may also contribute to spreads in radii, as its presence biases the SED fit towards higher inferred radii. 
In addition, we note that since we study the radius inflation effect in single dwarf stars, its applicability to evolved or interacting binary systems, such as RS Canum Venaticorum variables, may be limited. For example, in such systems, the activity is seen to be maximized in the most strongly interacting stars, and a direct analogue in the single star case is not immediately obvious \citep{2025arXiv250405561D}.

The table of derived parameters for our cluster sample is in Table \ref{tab:clusterdata}.

\begin{table}[!thb]
\centering
\caption{Column descriptions for the radius inflation cluster table.} \label{tab:clusterdata}
\begin{tabular*}{\columnwidth}{@{}l@{\hspace*{5pt}}l@{}}
  \hline
  Label & Contents \\
  \hline
  APOGEE\_ID & Source 2MASS ID \\
  RA\_J2000 & Right Ascension \\
  Dec\_J2000 & Declination \\
  Prot & Rotation period (d) \\
  Ro & Rossby number \\
  Radius & Stellar radius from SED fit (R$_\odot$) \\
  Teff & Two-temperature effective temperature (K) \\
  fspot & Two-temperature starspot filling fraction \\
  xspot & Two-temperature starspot temp. contrast \\
  logg & Two-temperature surface gravity \\
  vsini & Two-temperature rotational velocity (km/s) \\
  {[}M/H{]} & Two-temperature metallicity \\
  vdop & Two-temperature microturbulence (km/s) \\
  Av & Extinction from SED fit (mag) \\
  Fbol & Bolometric flux from SED fit \\
  e\_Radius & Uncertainty in derived radius \\
  e\_Teff & Uncertainty in derived Teff \\
  e\_fspot & Uncertainty in derived fspot \\
  e\_xspot & Uncertainty in derived xspot \\
  e\_logg & Uncertainty in derived logg \\
  e\_{[}M/H{]} & Uncertainty in derived [M/H] \\
  e\_vsini & Uncertainty in derived vsini \\
  e\_vdop & Uncertainty in derived vdop \\
  e\_Avm & Lower uncertainty range in derived extinction \\
  e\_Avp & Upper uncertainty range in derived extinction \\
  e\_Fbolm & Lower uncertainty range in derived bolometric flux \\
  e\_Fbolp & Upper uncertainty range in derived bolometric flux \\
  Flag\_Binary & Total binarity flag \\
  Flag\_Binary\_Gaia\_RV & RV binarity flag from {\it Gaia} \\
  Flag\_Binary\_APOGEE\_RV & RV binarity flag from APOGEE \\
  Flag\_Binary\_Gaia\_RUWE & RUWE binarity flag from {\it Gaia} \\
  Flag\_Binary\_Gaia\_Multiple & Imaging binarity flag from {\it Gaia} \\
  chisq\_fit\_SED & $\chi^2$ statistic of the SED fit \\
  Cluster\_Name & Upper Sco, $\alpha$ Per, Pleiades, or Praesepe \\
\hline
\multicolumn{2}{l}{This table is entirely available in machine-readable form.}
\end{tabular*}
\end{table}

\section{MCMC fitting of radius inflation relationships}\label{appendix:mcmc}

\begin{figure*}
    \centering
    \includegraphics[width=0.8\linewidth,trim=0 0 0 0,clip]{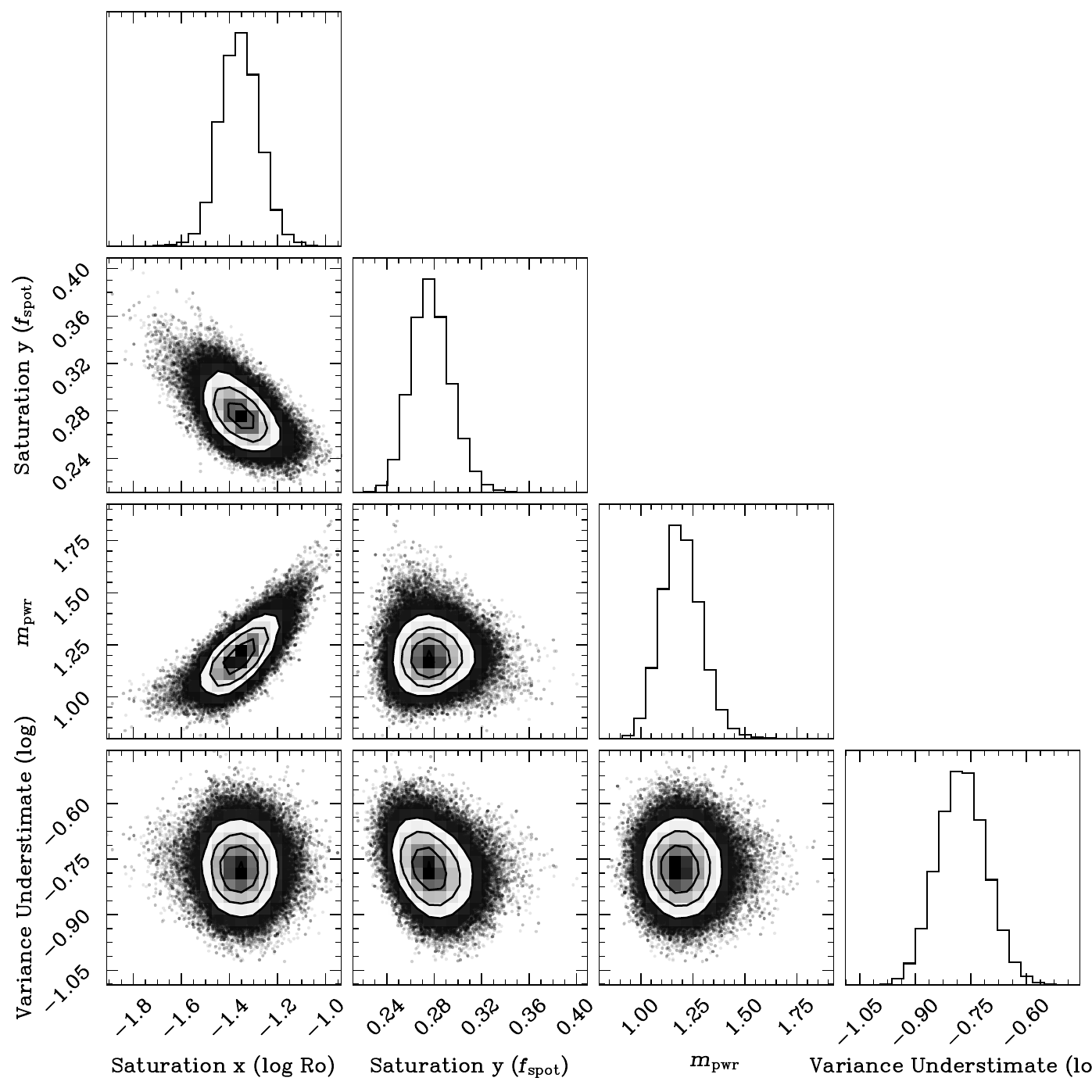}
    \caption{Markov Chain Monte Carlo corner plots showing parameters recovered for the power-law fit in Figure~\ref{fig:DelR_frac_vs_Ro}. The best-fit values are, for rotational saturation: $\log \mathrm{Ro}_{\mathrm{crit}} = -1.396^{+0.076}_{-0.070}$; spot saturation level: $f_{\mathrm{spot, \; sat}} = 0.276^{+0.017}_{-0.016}$; slope at $\log \mathrm{Ro} > \log \mathrm{Ro}_{\mathrm{crit}}$: $m_{\mathrm{slope}} = 1.154^{+0.087}_{-0.081}$.}
    \label{fig:corner}
\end{figure*}

We use \texttt{emcee} \citep{2013PASP..125..306F} to fit a power law to the radius inflation---stellar Rossby number plot in Figure~\ref{fig:corner}. We incorporate a parameter $\log f$, defined as the factor by which the variance is underestimated. The subsequent best-fit relation is expressed in Figure~\ref{fig:DelR_frac_vs_Ro}, with the standard deviation of the cluster points represented as the blue band, which has a width of $\pm 8.3$\%. Our fit for the saturation of the radius inflation signal implies that the saturated stars are around 28\% larger than expected by non-magnetic isochrones, before declining by a factor of 5 to a value of 5.5\% at a Rossby number of unity.

\begin{figure*}
    \centering
    \includegraphics[width=0.8\linewidth,trim=0 0 0 0,clip]{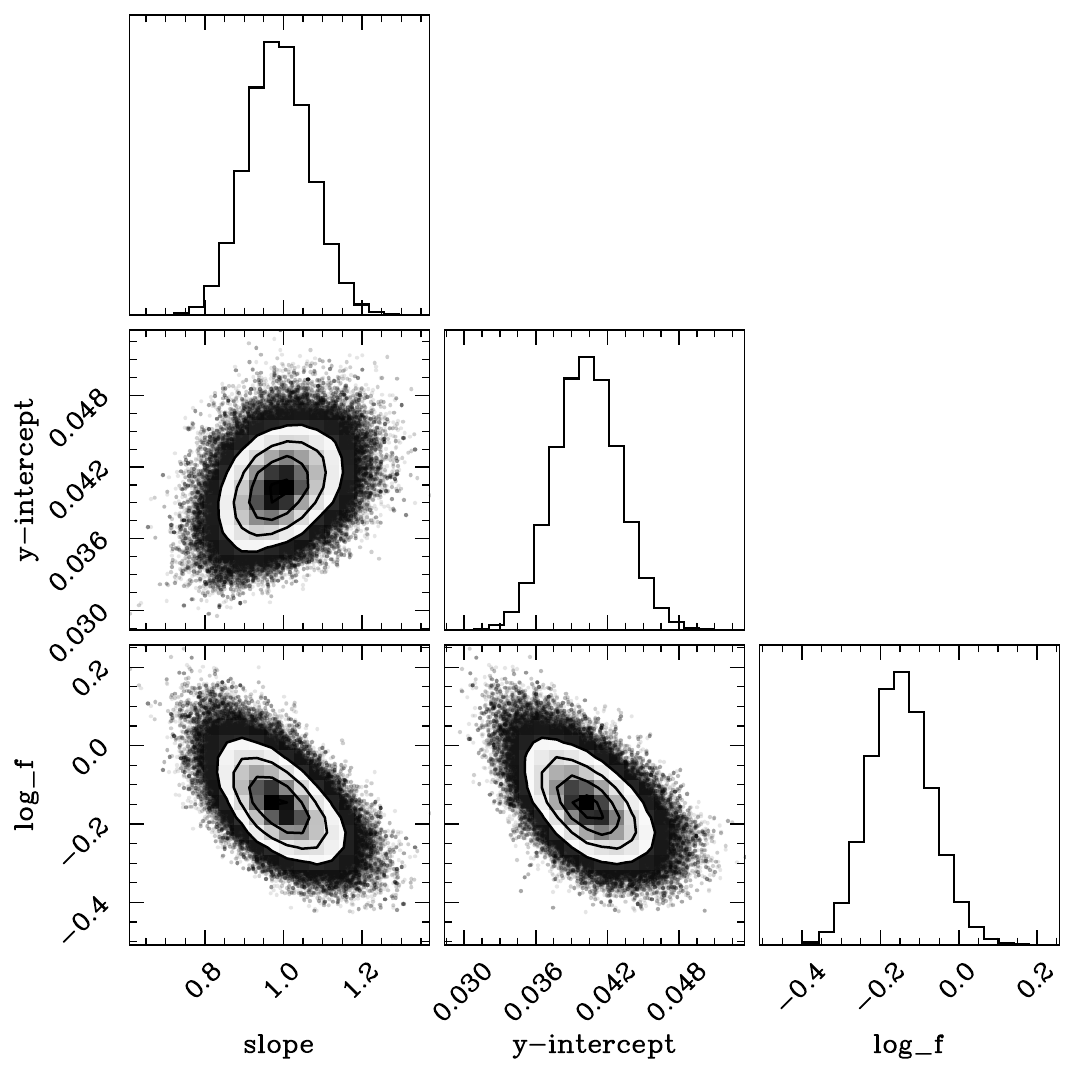}
    \caption{Markov Chain Monte Carlo corner plots showing parameters recovered for the linear fit in Figure~\ref{fig:DelR_obs_vs_spots}. The best-fit values are described in Section~\ref{sec:driver}.}
    \label{fig:cornerspots}
\end{figure*}

For the MCMC fit to Figure~\ref{fig:DelR_obs_vs_spots}, we show the corner plot in Figure~\ref{fig:cornerspots}, again using the parameter $\log f$ as defined in the prior fit. The best-fit relation is expressed in Figure~\ref{fig:DelR_obs_vs_spots}, with the standard deviation of the cluster points represented as the light gray band, which has a width of $\pm 0.086$~R$_\odot$.

\end{document}